\documentclass{elsart1p}
\usepackage{graphicx}
\usepackage{amssymb}
\newcommand{ \be }{\begin{eqnarray}}
\newcommand{ \ee }{\end{eqnarray}}
\newcommand{ \ben }{\begin{enumerate}}
\newcommand{ \een }{\end{enumerate}}
\newcommand{ \la }{\langle}
\newcommand{ \ra }{\rangle}

\newcommand{ \eps }{\varepsilon}

\usepackage{color}

\newcommand{\mysection}[1]{\vspace*{2mm} \noindent {\em #1} \vspace{1mm}}

\begin{document}
\begin{frontmatter}
%
%
%
\title{Anisotropic collective phenomena in ultra-relativistic 
nuclear collisions}
%
%
\author{Sergei A. Voloshin}
\address{Wayne State University, 666 W. Hancock, 48201 Detroit, MI, U.S.A.}
\begin{abstract}
For a detailed review of this subject I refer to a recent 
paper~\cite{Voloshin:2008dg}; in this talk I only very briefly comment on  
a few most important questions:
(a) Very recent significant progress in viscous hydrodynamics calculations
(b) Initial eccentricity/flow fluctuations, the effect of which has been
clarified recently
(c) Initial conditions, in particular the role of
 the gradients in the initial velocity field,
(d) Puzzling system size dependence of directed flow
(e) Azimuthal correlations that are sensitive to the strong parity
violation
(f) Future measurements at RHIC and LHC, including pp-collisions
\end{abstract}
\end{frontmatter}
%
%
%
%
\mysection{Introduction}

Anisotropic flow for a several years remains one of the most important
measurements in the field of heavy ion collision.
Those were the key measurements~\cite{Ackermann:2000tr} 
for making a conclusion on creation of
the strongly interacting Quark Gluon Plasma (sQGP) at RHIC.  
The observation of the constituent quark number 
scaling~\cite{Voloshin:2002wa,Molnar:2003ff} in
elliptic flow at intermediate transverse momenta 
is a strong evidence for deconfinement.
Recently the progress in this field has been reviewed,  
including many technical details, 
in~\cite{Voloshin:2008dg}.

\begin{figure}[t]
  \includegraphics[width=0.99\textwidth]{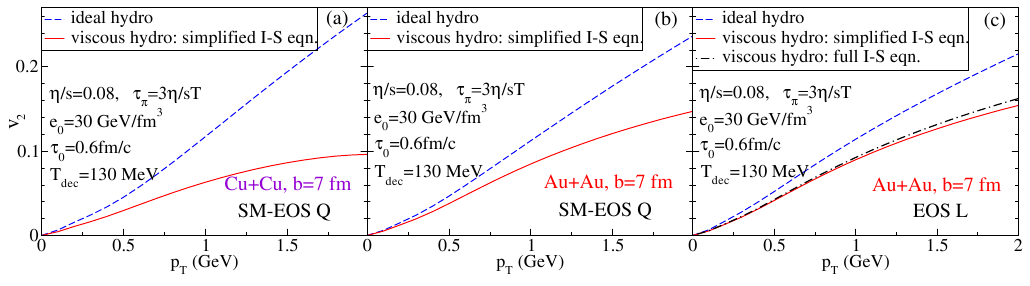}
\caption{Ideal and viscous hydro calculations for direct pions
  elliptic flow, with parameters as indicated~\cite{Song:2008si}}
\label{fig:song}
\end{figure}

\begin{figure}[t]
\begin{minipage}[t]{0.48\textwidth}
  \includegraphics[width=0.95\textwidth]{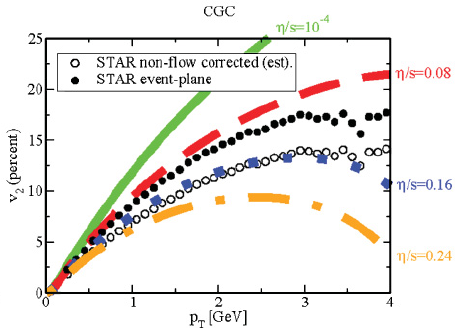}
\caption{Viscous hydro calculations~\cite{Luzum:2008cw}
compared to the STAR data.}
\label{fig:romat}
\end{minipage}
\hspace{0.03\textwidth}
\begin{minipage}[t]{0.48\textwidth}
  \includegraphics[width=0.95\textwidth]{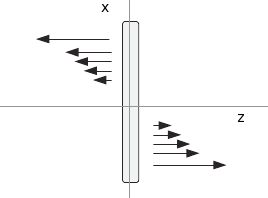}
\caption{Initial velocity profile in non-central nuclear 
collisions~\cite{Becattini:2007sr}}
\label{fig:becat}
\end{minipage}
\end{figure}

\mysection{Viscous hydrodynamics.}

The importance and the magnitude of {\em the viscous effects} could 
be judged already from the
early calculations~\cite{Teaney:2001av}
 where the hydro dynamical evolution at some intermediate stage was joined to
the transport model to simulate late (viscous) evolution of the
system.
Recently there have been performed several 
calculations of the hydrodynamical expansion with viscous terms
explicitly included into equations. 
A great advancement in these calculations (including  in the formulations
of the equations itself) has been achieved via the collaboration of several 
groups within TECHQM initiative~\cite{techqm}.
Even the ``minimal'' values of viscosity ($\eta/s=1/(4\pi)$) were
found to have a strong effect on elliptic flow (see,
Fig.~\ref{fig:song})
leading to up to $\sim$25-30\% reduction in flow
values in Au+Au collisions and as large as 50\% in Cu+Cu.
Such a strong sensitivity of the elliptic flow to viscosity values
can be used for measuring viscosity.
One of such attempts is presented in Fig.~\ref{fig:romat}, where the
calculations~\cite{Luzum:2008cw} 
at different viscosity values are compared to the STAR
data assuming CGC initial conditions. At present, the initial
conditions, and to somewhat lesser extend, the uncertainties in the
hadronization stage are the main factors preventing precise
measurement of viscosity. At the same time one can safely put an upper
bound on $\eta/s$ of about factor of five the minimal value 
of $1/(4\pi)$~\cite{Song:2008si}.

\mysection{Flow and eccentricity fluctuations}

The role of flow fluctuations and non-flow effects is one of a 
long standing problem that received a lot of attention and  
significant progress has been made in the recent
couple years. In particular, the role of fluctuations
in  the initial system geometry  defined by nuclear 
{\em participants}
(interacting nucleons or quarks) has been greatly
clarified~\cite{Bhalerao:2006tp,Voloshin:2007pc,Alver:2008zz}. 
At fixed impact parameter, the geometry of the {\em participant zone}
fluctuates, both, in terms of the value of the
eccentricity as well as the orientation of the major
axes. 
The anisotropy develops along the plane spanned by the minor
axis of the participant zone and the beam direction, 
the so called {\em participant plane}. As the true
reaction plane (defined by the impact parameter)
is not known and the event plane is estimated from
the particle azimuthal distribution ``defined'' by the participant plane, 
the apparent (participant plane) flow appears to be  always
bigger (and always ``in-plane'', $v_{2,PP}>0$) compared to the ``true''
flow as projected onto the reaction plane.  

It was noticed~\cite{Voloshin:2007pc} that in collisions of heavy nuclei the
fluctuations in the eccentricity 
$(\eps_x,\eps_y)=(\la (\sigma_y^2-\sigma_x^2)/(\sigma_y^2+\sigma_x^2)\ra,
\la (2\sigma_{xy}^2/(\sigma_y^2+\sigma_x^2)\ra)$ 
can be well described by two-dimensional Gaussian, for which 
the higher cumulant flow ($v\{n\}, n\ge 4$) is not only 
insensitive to non-flow but also to fluctuations. 
All of higher cumulants are
exactly equal to the ``true'' flow, namely as given by projection onto
the reaction plane. 
This greatly simplifies the comparison of theoretical calculations
to the data, as it says that in such calculation one should not
worry how to take into account the fluctuations in the initial
eccentricity (which is a non-trivial task) but just compare to the
``right'' measurement, e.g. $v_2\{4\}$.  
At the same time, the apparent (participant plane)
flow become unmeasurable in a sense that flow fluctuations could not
be separated from non-flow contributions by means of correlation
measurements.       

The role of fluctuations and non-flow in the 
{\em event plane method} is more complicated to investigate due to
non-linearity of the problem. Nevertheless, first with 
Monte-Carlo~\cite{Alver:2008zz}
and later analytically~\cite{Olli} 
in small fluctuation limit this problem also
has been solved. It appears that this method also does not allow
to separate two effects.

Now we  have almost full understanding of the role of
fluctuations and non-flow in different flow measurements.
Unfortunately this progress in understanding the nature of fluctuations
does not help in resolving the problem of measuring separately 
flow fluctuations and non-flow. One needs further assumptions, e.g. as
done by the PHOBOS Collaboration that 
uses estimates of non-flow effects from correlations
with large rapidity separations.

\mysection{ Initial flow velocity profile. Elliptic and Directed flow.}

Another important and interesting direction that is just started 
to be explored is the role of the non-zero initial flow velocity
profile, e.g. non-zero
velocity  gradient along the impact parameter, Fig.~\ref{fig:becat}. 
As shown in~\cite{Becattini:2007sr} such a gradient directly contributes to the
in-plane expansion rate (see Eq.~22 in~\cite{Becattini:2007sr}). 
The contribution to the final magnitude of the elliptic flow can be
significant; to check this we need full 3d hydrodynamics study with
different initial conditions.

Note that such initial flow gradients naturally
would lead to {\em directed} flow (see the same Eq.~22); 
this question was briefly
addressed in~\cite{Troshin:2007cp}. 
Speculating on this subject one would notice that viscous effects
must also play an important role in such a scenario. 
It will be very interesting to compare the calculations in such a model to 
a very precise recent data from STAR~\cite{Abelev:2008jga}.

One surprising observation made in~\cite{Abelev:2008jga} is that
the directed flow is almost independent of the system size if compared
at the same centrality. 
It is not described by any model.
Understanding of such behavior can be very
important in clarification of the initial conditions.
Recall that predictions for
non-trivial dependence of directed flow on
rapidity~\cite{Snellings:1999bt}  (so called
``wiggle'') was based on the assumption of non-zero initial velocity profile of
net nucleons similar to that shown in Fig.~\ref{fig:becat}.
Future measurements of directed flow with identified particles will be
very important in this respect. Another possibility to address this
question would be colliding beams of nuclei of the same mass but
different charge, similar to what has been done at GSI at lower
energies, where beams of $^{96}_{44}Ru$ and    $^{96}_{40}Zr$ were 
used~\cite{Hong:2001tm,Hong:2003jk}.
As discussed below such isobaric beams will be very also important for the
search of strong parity violations.

\begin{figure}[t]
\begin{minipage}[t]{0.49\textwidth}
  \includegraphics[width=0.95\textwidth]{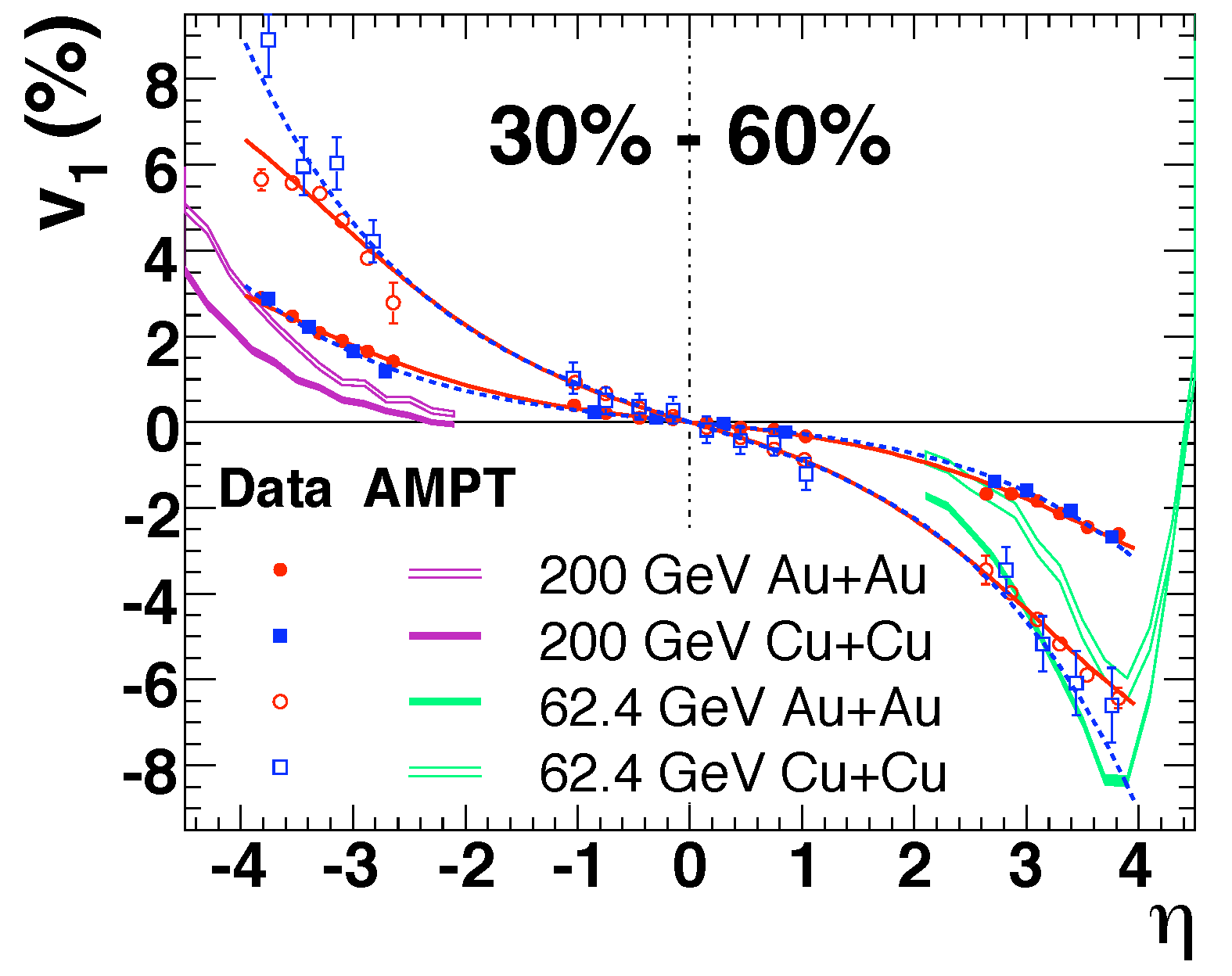}
\caption{Directed flow in Au+Au and Cu+Cu collisions as measured by the
STAR Collaboration~\cite{Abelev:2008jga}}
\label{fig:v1}
\end{minipage}
\hspace{0.01\textwidth}
\begin{minipage}[t]{0.48\textwidth}
  \includegraphics[width=0.95\textwidth]{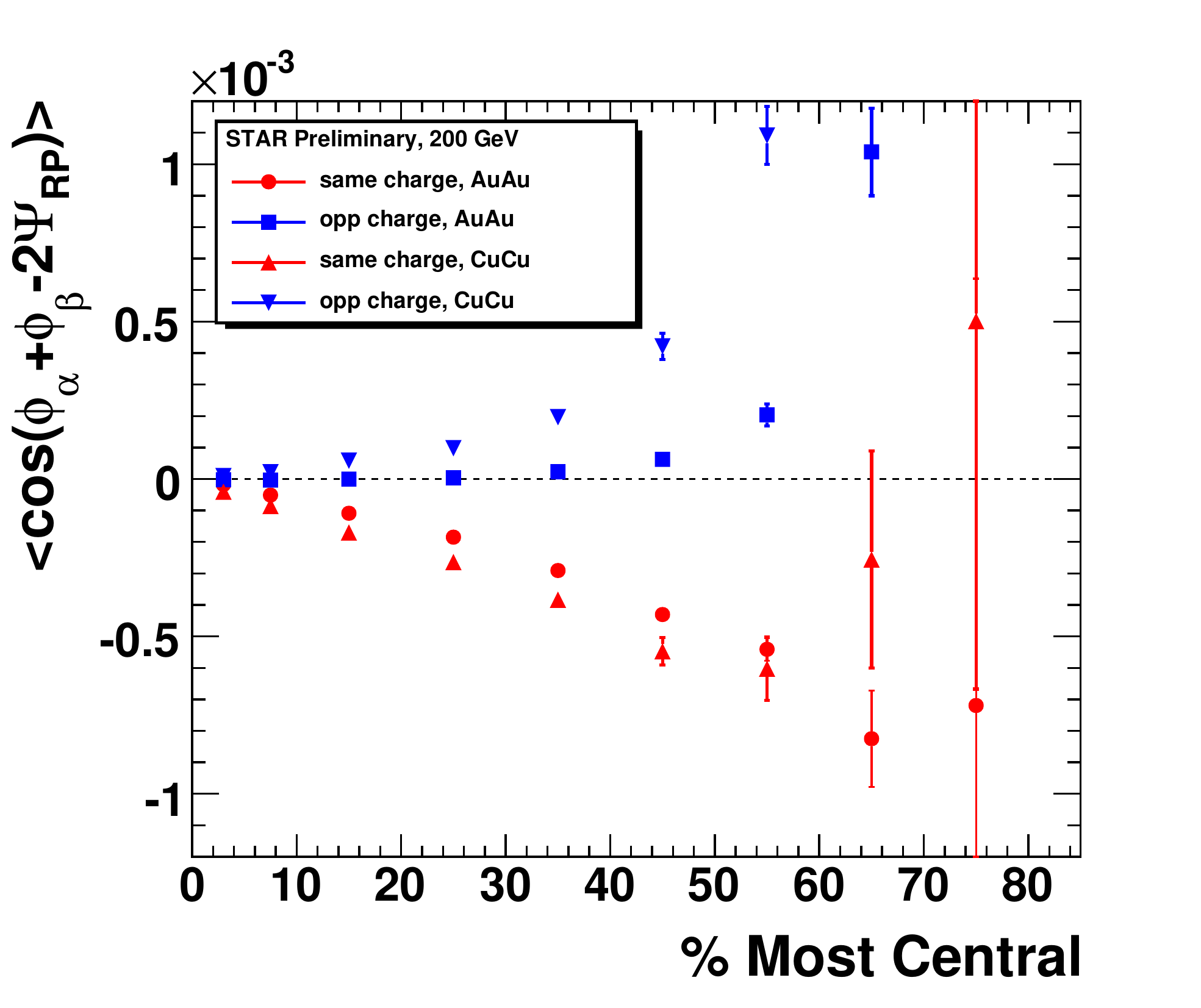}
\caption{Preliminary results~\cite{Voloshin:2004vk}
 on azimuthal anisotropy correlator sensitive to the strong 
$\cal P-$violation effects.}
\label{fig:parity}
\end{minipage}
\end{figure}

\mysection{Search for the strong $\cal P-$violation}

It was shown in~\cite{Kharzeev:2004ey,Fukushima:2008xe} that
in the presence of topologically non-trivial gluonic fields the 
magnetic field of the colliding nuclei induces parallel to it
electric field (the effect which violates parity). 
The induced electric filed leads to the charge separation 
(preferential emission of same
charge particles)  in the direction perpendicular to
the reaction plane. 
Such anisotropy, which very much resembles
``out-of-plane directed flow'' can be addressed with the
help of three-particle correlations~\cite{Voloshin:2004vk}
by measuring $\la \cos(\phi_\alpha+\phi_\beta-2\Psi_{RP}) \ra$,
where $\phi_{\alpha,\beta}$ are azimuthal angles of two (same or
 opposite) charged particles, and $\Psi_{RP}$ is the reaction plane angle. 
The estimates~\cite{Kharzeev:2007jp}
indicate that the effect is
 strong enough to be observed in heavy ion collisions.
The STAR Collaboration reported
the preliminary results~\cite{Voloshin:2004vk},
see Fig.~\ref{fig:parity}, that qualitatively agree with
theoretical estimates~\cite{Kharzeev:2004ey,Kharzeev:2007jp}. 
Note, that the correlator 
$\la \cos(\phi_\alpha+\phi_\beta-2\Psi_{RP}) \ra$  is $\cal P$-even and
contain contributions from other effects not related to parity
violation. 
A careful analysis of such a contribution is obviously
needed before any strong conclusion can be drawn from these measurements. 

Taking into account the importance of the question,
one can envision a dedicated program for establishing the
nature of the signal and further detail study. 
From the theoretical point of view, the calculation of the dependence
on centrality
and system system size looks fully doable though requires
significant computing and man power (e.g. 3d hydrodynamics is needed for
the calculation of the magnetic field).
Detailed predictions on the transverse momentum and particle type
dependence also will be essential in differentiating this effect from
possible ``background'' contributions. 
Also interesting would be a calculations of ``usual'' transverse
momentum and rapidity two and
multiparticle correlations
due to topological effects responsible for the charge separation.
There can be extensive experimental program. 
For example, the energy dependence (e.g. during the RHIC beam energy scan)
of the effect can be very indicative 
if any threshold type behavior
will be found, as the effect might be strongly suppressed in no QGP systems. 
Identified and multiparticle correlations studies also
will be available with larger statistics.   
The charge separation dependence on the magnetic
field~\cite{Kharzeev:2007jp} can be tested with collision of isobaric
nuclei, such as  $^{96}_{44}Ru$ and    $^{96}_{40}Zr$ that were 
used at GSI~\cite{Hong:2001tm,Hong:2003jk} and
discussed above in relation to the directed flow studies 
(in this case one needs symmetric collisions).

\mysection{RHIC: beam energy scan. LHC: Pb+Pb and p+p.}

Coming years promise many new interesting data from low energy RHIC run
and, of course, from LHC. The main interest in the low energy RHIC scan, 
anisotropic flow is no exception, is the search for the
QCD critical point. The scan would cover the energy region from top
AGS energies, over the CERN SPS, and higher. In terms of
anisotropic flow two major observables to watch would be a possible
``wiggle'' in $v_2/\eps$ dependence on particle density~\cite{Voloshin:1999gs}
 and
``collapse'' of directed flow~\cite{Stocker:2007pd}.
 RHIC also has plans to extend its reach
in terms of energy density using uranium beams. From the first
estimates and ideas of using uranium beam we now have real detailed
simulations~\cite{Nepali:2007an} of such collisions 
with developed methods for a selection of
desired geometry of the collision. 

The predictions for the LHC are rather uncertain, though most agree
that the elliptic flow will continue to increase~\cite{Abreu:2007kv},
partially due to smaller viscous effects.  
Simple extrapolations~\cite{Busza:2007ke,Borghini:2007ub},
of the $v_2$ collision energy dependence to LHC energies
 lead to about 20-30\% increase in elliptic flow values.
Note that there exist calculations predicting {\em decrease} of the
elliptic flow~\cite{Krieg:2007sx}. 
Another important observation is an increase in mass dependence
(splitting) of $v_2(p_t)$ due to a strong increase of radial flow.

\begin{figure}[t]
\begin{minipage}[t]{0.65\textwidth}
  \includegraphics[width=0.9\textwidth,height=0.75\textwidth]{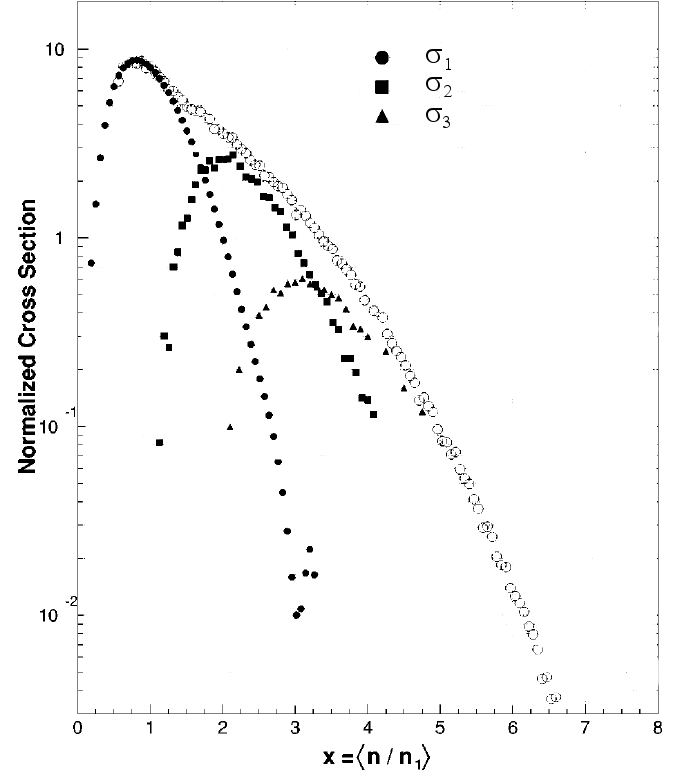}
\caption{Multiplicity distribution, $\sqrt{s}=1.8$~TeV, as a
  superposition of events with different number of soft parton 
collisions~\cite{Walker:2004tx}} 
\label{fig:kno}
\end{minipage}
\begin{minipage}[t]{0.35\textwidth}
 \centerline{ \includegraphics[width=0.9\textwidth]{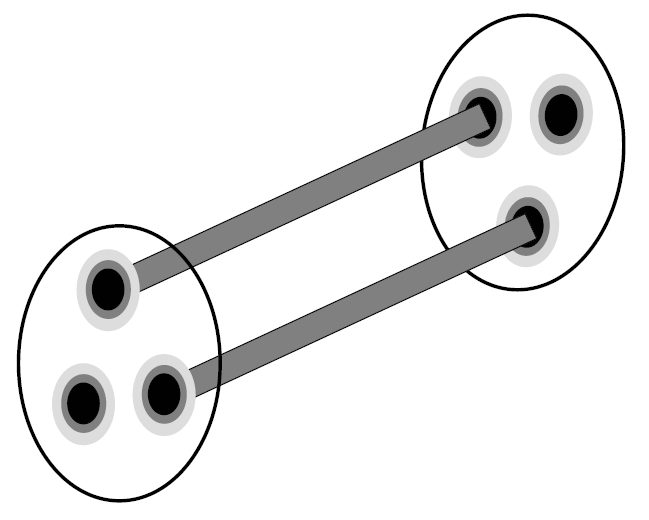}}
\caption{Two ``string'' pp event.}
\label{fig:ppq}
\end{minipage}
\end{figure}

An exciting direction at LHC will be the study of collective effects in
pp collisions. Note that event multiplicities at LHC energies
will be comparable to those of central Cu+Cu collisions at RHIC.
The detailed analysis of event anisotropies will be possible and very
interesting; it promises new insights into physics of multiparticle
production. I mention here only one possibility - the study of the 
so-called multi parton collisions. Fig.~\ref{fig:kno} shows the
multiplicity distribution measured by E835 Collaboration at Fermilab
in the so called KNO variable. It is decomposed~\cite{Walker:2004tx} 
into distributions corresponding to events with one, two or three soft
parton interactions. The nature and space-time picture of these
interaction is not totally clear. One possibility would be that it
corresponds to interactions of different number of constituent
quarks. Fig.~\ref{fig:ppq} shows a schematic view of such
an interaction. Experimentally this question can be addressed
 by studying azimuthal
multiplicity and transverse momentum correlations as a function of
total event multiplicity.  

\vspace*{1mm}
In summary, we have had very exciting years of anisotropic phenomena study,
which greatly enriched our understanding of ultra-relativistic nuclear
collisions and multiparticle production in general. Future
experimental programs at LHC and RHIC promise new results and new physics.

\vspace*{1mm}
I thank the Organizers for the invitation  and
A. Poskanzer for numerous fruitful discussions.

\vspace*{3mm}

\end{document}